\documentclass{emulateapj}
\usepackage{lscape}

\newcommand{\kms}{\ifmmode{\rm km\thinspace s^{-1}}\else km\thinspace s$^{-1}$\fi}

\shortauthors{Torres \& Lacy}
\shorttitle{VZ~Cep}

\begin{document}

\journalinfo{Accepted for publication in The Astronomical Journal}

\title{Absolute dimensions of the F-type eclipsing binary star
VZ~Cephei}

\author{
Guillermo Torres\altaffilmark{1}, and
Claud H.\ Sandberg Lacy\altaffilmark{2}
}

\altaffiltext{1}{Harvard-Smithsonian Center for Astrophysics, 60
Garden Street, Cambridge, MA 02138, e-mail: gtorres@cfa.harvard.edu}

\altaffiltext{2}{Department of Physics, University of Arkansas,
Fayetteville, AR 72701, e-mail: clacy@uark.edu}

\begin{abstract}

We present new $V$-band differential photometry and radial-velocity
measurements of the unevolved 1.18-day period F+G-type double-lined
eclipsing binary VZ~Cep. We determine accurate values for the absolute
masses, radii, and effective temperatures as follows: $M_{\rm A} =
1.402 \pm 0.015$~M$_{\sun}$, $R_{\rm A} = 1.534 \pm 0.012$~R$_{\sun}$,
$T_{\rm eff} = 6690 \pm 160$~K for the primary, and $M_{\rm B} =
1.1077 \pm 0.0083$~M$_{\sun}$, $R_{\rm B} = 1.042 \pm
0.039$~R$_{\sun}$, $T_{\rm eff} = 5720 \pm 120$~K for the secondary. A
comparison with current stellar evolution models suggests an age of
1.4~Gyr for a metallicity near solar. The temperature difference
between the stars, which is much better determined than the absolute
values, is found to be $\sim$250~K larger than predicted by theory. If
all of this discrepancy is attributed to the secondary (which would
then be too cool compared to models), the effect would be consistent
with similar differences found for other low-mass stars, generally
believed to be associated with chromospheric activity. However, the
radius of VZ~Cep~B (which unlike the primary, still has a thin
convective envelope) appears normal, whereas in other stars affected
by activity the radius is systematically larger than predicted. Thus,
VZ~Cep poses a challenge not only to standard theory but to our
understanding of the discrepancies in other low-mass systems as well.

\end{abstract}


\section{Introduction}

VZ~Cephei (also known as BD~+70~1199 and GSC~04470-01334; $\alpha =
21^{\rm h}\,50^{\rm m}\,11\fs14$, $\delta =
+71\arcdeg\,26\arcmin\,38\farcs3$, J2000.0; $V = 9.72$) was discovered
photographically as a variable star by \cite{Gengler:28}, who
classified it to be of type ``Is?'', a rapid irregular variable.
\cite{Cannon:34} made the first spectral type assignment as G0.  Its
discovery as an eclipsing binary is due to \cite{Rossiger:78}, who
determined a period of 1.18336 days and showed it to have unequal
minima.  The system was included by \cite{Lacy:92, Lacy:02a} in his
photometric surveys of eclipsing binary stars.  Because of its
relatively late spectral class, \cite{Popper:96} had it as a target in
his program to search for late-type (F--K) eclipsing binary stars.  He
concluded on the basis of 4 spectrograms that both stellar components
were likely hotter than G0.  No determinations have been made of the
physical properties of the stars, and the system has generally been
neglected except for measurements of the times of eclipse made by a
number of investigators since 1994.

We began our investigation for the same reason Popper did: to test
theoretical predictions of the properties of low-mass stars.  We and
other authors have previously found that in some of these binary
systems the absolute properties are not well described by standard
stellar evolution theory \citep[see, e.g.,][]{Popper:97, Clausen:99,
Ribas:06, Torres:06}.  We find below that VZ~Cep also shows some
anomalies compared to standard models, even though its components are
both more massive than the Sun.

\section{Eclipse ephemeris}
\label{sec:ephemeris}

Photometric times of minimum light of VZ~Cep available from the
literature are collected in Table~\ref{tab:minima}. Eclipse
ephemerides determined by weighted least squares separately from the
15 primary minima and the 13 secondary minima gave the same period
within the errors. A joint fit of all the measurements was then
performed enforcing a common period but allowing the primary and
secondary epochs to be free parameters.  Scale factors for the
internal errors were determined by iterations separately for the two
types of measurements in order to achieve reduced $\chi^2$ values near
unity. This solution resulted in a phase difference between the two
epochs of $\Delta\phi = 0.50033 \pm 0.00022$, not significantly
different from 0.5.  For our final ephemeris we imposed a circular
orbit, and obtained
\begin{eqnarray*}
{\rm Min~I~(HJD)} & = & 2,\!452,\!277.324478(59) + 1.183363762(84)\cdot E .
\end{eqnarray*}
The uncertainties of the fitted quantities in terms of the least
significant digit are shown in parentheses. The final scale factors
for the published internal errors were similar to those in the
previous fit, and were 1.38 for the primary and 2.42 for the
secondary.  We adopt this ephemeris for the spectroscopic and
photometric analyses below.

\section{Spectroscopic observations and orbit}
\label{sec:spectroscopy}

VZ~Cep was placed on the observing list at the Harvard-Smithsonian
Center for Astrophysics (CfA) in 2003 January, and was observed until
2007 June with an echelle spectrograph on the 1.5-m Tillinghast
reflector at the F.\ L.\ Whipple Observatory (Mount Hopkins, AZ). A
single echelle order 45\,\AA\ wide centered at 5188.5\,\AA\ was
recorded with an intensified Reticon photon-counting diode array, at a
resolving power of $\Delta\lambda/\lambda \approx 35,\!000$. The
strongest lines in this window are those of the \ion{Mg}{1}~$b$
triplet. A total of 39 spectra were obtained with signal-to-noise
ratios ranging from 19 to 47 per resolution element of 8.5~\kms.

Radial velocities were measured with the two-dimensional
cross-correlation technique TODCOR \citep{Zucker:94}. Templates for
the primary and secondary were selected from a large library of
synthetic spectra based on model atmospheres by R.\ L.\
Kurucz\footnote{Available at {\tt http://cfaku5.cfa.harvard.edu}.}
\citep[see also][]{Nordstrom:94, Latham:02}. These calculated spectra
cover a wide range of effective temperatures ($T_{\rm eff}$),
rotational broadenings ($v \sin i$ when seen in projection), surface
gravities ($\log g$), and metallicities. Solar metallicity was assumed
throughout, along with initial values of $\log g = 4.5$ for both
components.  The temperatures and rotational velocities for the
templates were determined by running extensive grids of
two-dimensional cross-correlations and seeking the best correlation
value averaged over all exposures, as described in more detail by
\cite{Torres:02}. The secondary component in VZ~Cep is some 5 times
fainter than the primary, and we were unable to determine its
temperature independently. We therefore adopted the results from other
estimates described below, and chose the nearest value in our library,
which is 5750~K. Due to the narrow wavelength range of our spectra the
derived temperatures are strongly correlated with the assumed surface
gravities. The secondary $\log g$ presented in
\S\,\ref{sec:dimensions} is very close to the value we assumed, but
the primary $\log g$ is intermediate between 4.0 and 4.5, so we
repeated the calculations above using the lower value, and
interpolated. The results for the primary are $T_{\rm eff} = 6690 \pm
150$~K and $v \sin i = 57 \pm 3$~\kms, and for the secondary we obtain
$v \sin i = 50 \pm 10$~\kms. Radial velocities were derived with
template parameters near these values.  The stability of the
zero-point of the CfA velocity system was monitored by means of
exposures of the dusk and dawn sky, and small run-to-run corrections
were applied in the manner described by \citet{Latham:92}.

\begin{figure}
\epsscale{1.35}
\vskip -0.5in
{\hskip -0.25in \plotone{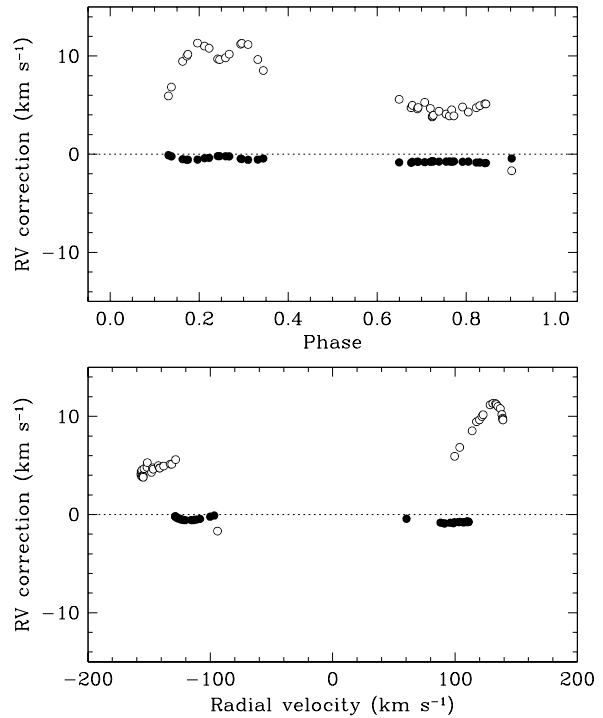}}
\vskip -0.7in
\figcaption[]{Corrections for systematics in the radial-velocity
measurements for VZ~Cep as a function of orbital phase and radial
velocity (see text).  Filled circles correspond to the primary and
open circles to the secondary.
\label{fig:rvcorr}}
\end{figure}

Possible systematics in the radial velocities that may result from
residual line blending in our narrow spectral window, or from shifts
of the spectral lines in and out of this window as a function of
orbital phase, were investigated by performing numerical simulations
as described by \cite{Torres:97, Torres:00}. Briefly, we generated
artificial composite spectra by adding together copies of the two
templates with scale factors in accordance with the light ratio
reported below, and with Doppler shifts for each star appropriate for
each actual time of observation, computed from a preliminary orbital
solution. These simulated spectra were then processed with TODCOR in
the same manner as the real spectra, and the input and output
velocities were compared.  Experience has shown that the magnitude of
these effects is difficult to predict, and must be studied on a
case-by-case basis. Corrections were determined for VZ~Cep based on
these simulations and were applied to the raw velocities. The
corrections for the primary star are small (under 1~\kms), but for the
secondary they are as large as 11~\kms, and as expected they vary
systematically with orbital phase or radial velocity (see
Figure~\ref{fig:rvcorr}). Similarly large corrections have been found
occasionally for other systems using the same instrumentation
\citep[e.g., AD~Boo, GX~Gem;][]{Clausen:08, Lacy:08}. The impact of
these corrections is quite significant for VZ~Cep: the minimum masses
increase by 4\% for the primary and 1.9\% for the secondary, and the
mass ratio decreases by 2.1\%.  The final velocities in the
heliocentric frame, including the corrections for systematics, are
listed in Table~\ref{tab:rvs} and have typical uncertainties of
1.3~\kms\ for the primary and 3.8~\kms\ for the fainter secondary.

\begin{figure}
\epsscale{1.35}
\vskip -0.1in
{\hskip -0.2in \plotone{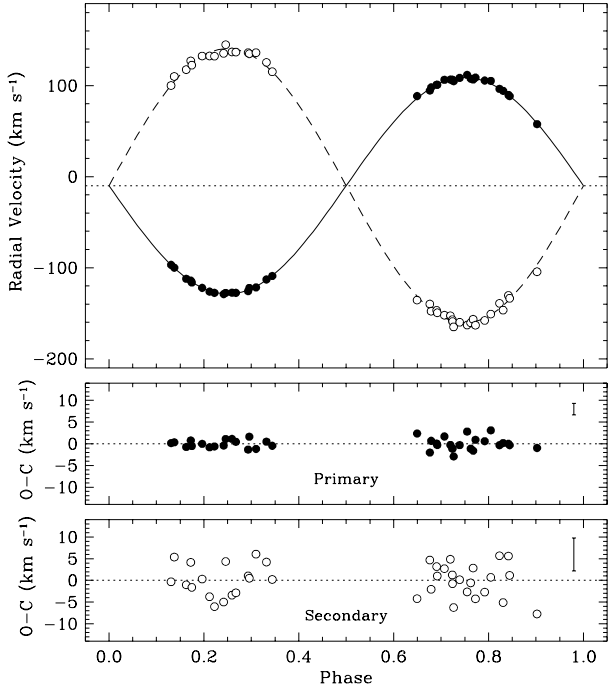}}
\vskip -0.1in
\figcaption[]{Radial-velocity measurements for VZ~Cep (including the
corrections for systematics described in the text) along with our
spectroscopic orbital solution. Filled circles correspond to the
primary, and the dotted line represents the center-of-mass
velocity. Error bars are smaller than the size of the points. The
$O\!-\!C$ residuals are shown on an expanded scale in the bottom
panels, where typical error bars are indicated in the upper right
corner.
\label{fig:rvorbit}}
\end{figure}

Preliminary single-lined orbital solutions using the primary and
secondary velocities separately indicated a zero-point difference
between the two data sets (i.e., a difference in the systemic velocity
$\gamma$), which is often seen by many investigators in cases where
there is a slight mismatch between the templates used for the
cross-correlations and the spectra of the real stars \citep[see,
e.g.,][]{Popper:00, Griffin:00}.  Numerous tests with other templates
did not remove the offset.\footnote{As a further test, solutions
without applying the corrections for systematics described in the
preceding paragraph gave an offset more than twice as large.} This
most likely arises in our case because of stellar parameters
(particularly the rotation) that fall in between the template
parameters available in our library of synthetic spectra, which come
in rather coarse steps of 10~\kms\ at the high rotation rates of
VZ~Cep. We therefore included this velocity offset as an additional
free parameter in the double-lined orbital fit, and we verified that
when doing so the velocity semi-amplitudes (which determine the
masses) are insensitive to the exact template parameters within
reasonable limits, and are essentially identical to those resulting
from separate single-lined solutions.  Our final orbital fit is
presented in Table~\ref{tab:specorbit}.  No indication of eccentricity
was found, as expected for such a short period, so only a circular
orbit was considered in the following. A graphical representation of
the observations and our best fit, along with the residuals, is shown
in Figure~\ref{fig:rvorbit}.

The light ratio between the primary and secondary was determined from
our spectra following \cite{Zucker:94}, accounting for the difference
in line blocking between the primary and the much cooler secondary.
After corrections for systematics analogous to those described above,
we obtained $L_{\rm B}/L_{\rm A} = 0.19 \pm 0.02$ at the mean
wavelength of our observations. A further adjustment to the visual
band taking into account the temperature difference between VZ~Cep A
and B was determined from synthetic spectra integrated over the $V$
passband and the spectral window of our observations, and resulted in
$(L_{\rm B}/L_{\rm A})_V = 0.22 \pm 0.02$.

\section{Photometric observations and analysis}
\label{sec:photometry}

\begin{figure}
\epsscale{1.25}
\vskip -0.6in
{\hskip -0.1in \plotone{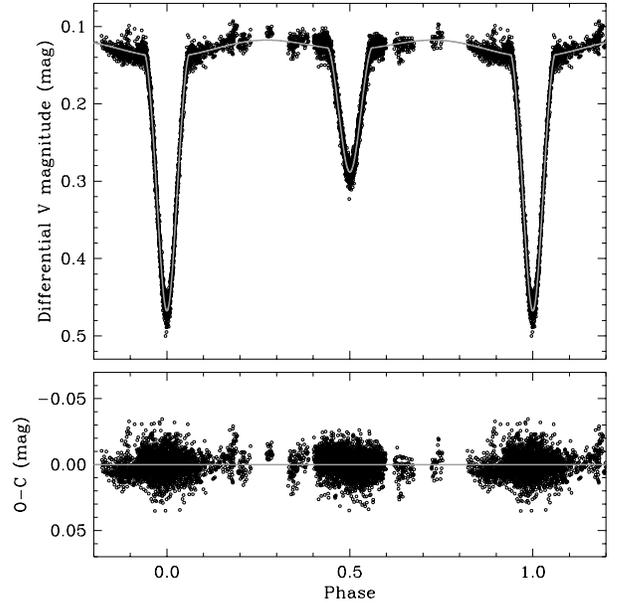}}
\vskip -0.75in
\figcaption{$V$-band photometric measurements for VZ~Cep, along with
our best constrained model described in the text. $O\!-\!C$ residuals
are shown at the bottom.\label{fig:photometry}}
\end{figure}

\begin{figure}
\epsscale{1.25}
\vskip -0.6in
{\hskip -0.1in \plotone{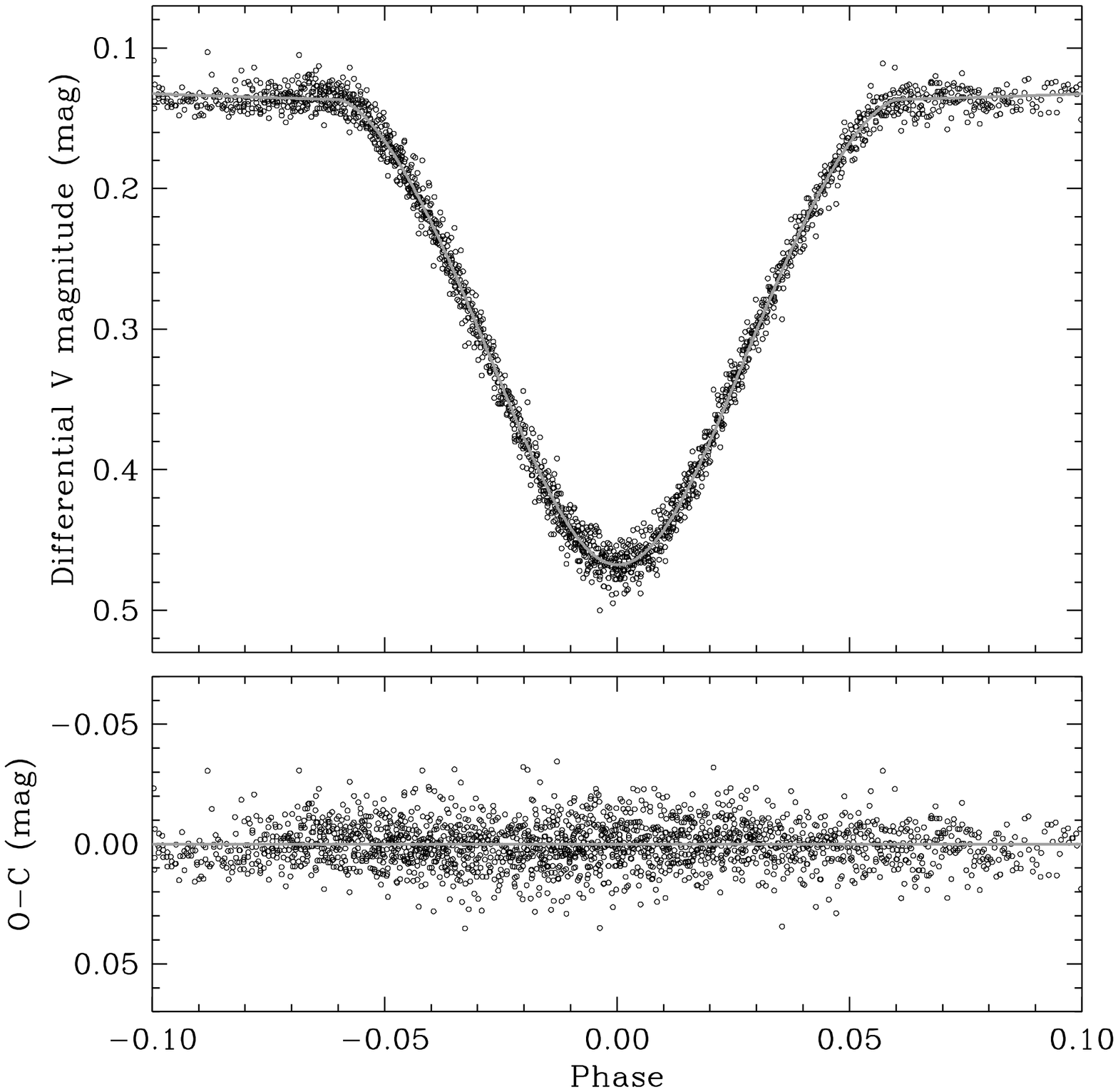}}
\vskip -0.75in
\figcaption{Enlarged view of Figure~\ref{fig:photometry} showing the
$V$-band photometry for VZ~Cep around the primary minimum.  $O\!-\!C$
residuals are shown at the bottom.\label{fig:photometry1}}
\end{figure}

\begin{figure}
\epsscale{1.25}
\vskip -0.6in
{\hskip -0.1in \plotone{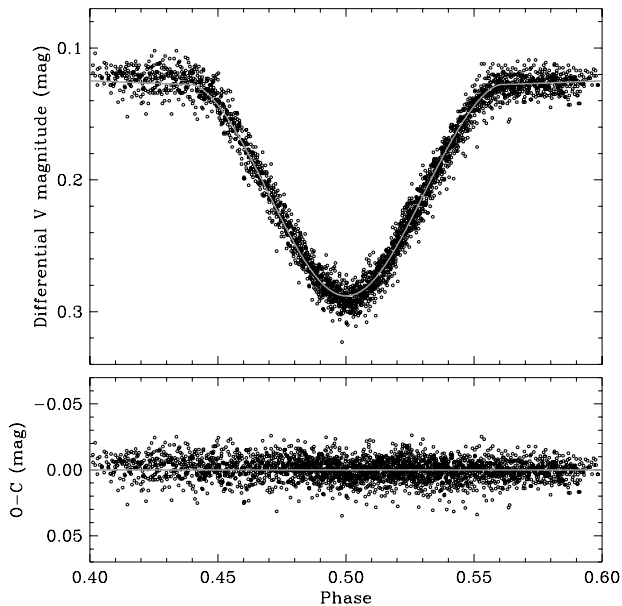}}
\vskip -0.75in
\figcaption{Enlarged view of Figure~\ref{fig:photometry} showing the
$V$-band photometry for VZ~Cep around the secondary minimum.
$O\!-\!C$ residuals are shown at the bottom.\label{fig:photometry2}}
\end{figure}

Differential photometry of VZ~Cep was obtained at the URSA Observatory
on the University of Arkansas campus at Fayetteville, AR. The URSA
Observatory sits atop Kimpel Hall and consists of a Meade f/6.3,
10-inch Schmidt-Cassegrain telescope with a Santa Barbara Instruments
Group ST8EN CCD camera inside a Technical Innovations RoboDome, all
controlled by a Macintosh G4 computer in an adjacent control room
inside the building. The field of view is 20$\arcmin\times30\arcmin$.
Images of VZ~Cep ($V \approx 9.7$) and the two comparison stars
GSC~04470-01497 ($V \approx 9.9$) and GSC~04470-01622 ($V \approx
11.2$), both of which are within 6\arcmin\ of the target, were taken
with typical integration times of 60 seconds through a Bessell $V$
filter.  With an overhead of about 30 seconds to download the images
from the camera, the observing cadence was typically 90 seconds per
image.  A ``virtual measuring engine'' application written by Lacy was
used to determine the brightness of the variable and comparison stars,
to subtract off the sky brightness, and to correct for differences in
airmass.  A total of 5473 images were gathered between 2001 March 5
and 2003 September 7. Differential magnitudes were formed between the
variable star and the magnitude corresponding to the sum of the fluxes
of the two comparison stars.  These are listed in
Table~\ref{tab:photometry}, and shown graphically in
Figure~\ref{fig:photometry} along with our modeling described
below. Expanded views of the primary and secondary eclipse are given
in Figure~\ref{fig:photometry1} and Figure~\ref{fig:photometry2}. The
typical precision of these measurements is about 0.007 mag, which is
comparable to that expected from photon statistics
($\sim$0.006~mag). The comparison stars are not known to be
variable. The mean magnitude difference between the two was constant
with a standard deviation of 0.0095~mag over 67 nights, which is what
would be expected for individual magnitudes with a typical error of
0.007~mag. A Lomb-Scargle periodogram analysis of the individual
differences was performed to search for periodic signals in either
star, but none were detected.

We have previously found \citep{Lacy:08} that the URSA photometry is
significantly improved by removal of small nightly zero-point
variations.  Thus 67 nightly corrections were made to the original
magnitudes based on a preliminary photometric orbital fit. This
procedure reduced the residual standard deviation by about 15\%, a
small but significant amount. Examination of these offsets, which are
typically smaller than 0.01~mag, revealed no detectable pattern as a
function of orbital phase. Such a pattern might be expected, for
instance, if there were perturbations in the light curve due to spots
on either star (assuming synchronous rotation). We thus consider the
nightly offsets to be instrumental in nature.

The corrected photometry was fitted with the NDE model
\citep{Etzel:81, Popper:81}, with all observations being assigned
equal weight. In this model the stars are represented as biaxial
ellipsoids, and despite the relatively large radius of the primary of
VZ~Cep relative to the separation (see below), its ellipticity of
0.016, as defined by \cite{Etzel:81}, is still well below the maximum
tolerance of 0.04 \citep{Popper:81}, and thus the model is expected to
be adequate for this case. We return to this below.  We used the
JKTEBOP implementation of \cite{Southworth:07} with a linear
limb-darkening law, consistent with our findings \citep{Lacy:05,
Lacy:08} that with the amount and precision of our data, non-linear
limb-darkening laws do not improve the accuracy of the fits
significantly.  The following quantities were allowed to vary in this
unconstrained solution: the central surface brightness $J_{\rm B}$ of
the secondary (cooler) star relative to the primary, the sum of
relative radii $r_{\rm A}+r_{\rm B}$, the ratio of radii $r_{\rm
B}/r_{\rm A}$, the orbital inclination $i$, the limb-darkening
coefficients $u_{\rm A}$ and $u_{\rm B}$, a phase offset, and the
magnitude at quadrature.  The following quantities were kept fixed:
the orbital eccentricity $e = 0$, the mass ratio from the
spectroscopic solution $q \equiv M_{\rm B}/M_{\rm A} = 0.7900$, and
the gravity brightening exponents $y_{\rm A} = 0.25$ and $y_{\rm B} =
0.36$, set by the temperatures and surface gravities following
\cite{Claret:98}.  The uncertainties of the adjustable parameters were
estimated with a Monte Carlo technique in which we generated 500
synthetic light curves, solved for the parameters, and calculated the
standard error of each parameter.  This process yielded uncertainty
estimates accurate to two significant digits, which is sufficient for
our purposes.  These ``Unconstrained'' results are given in
Table~\ref{tab:ebopfits}.  Tests allowing for non-zero eccentricity
and third light gave statistically insignificant values for those
quantities.

\begin{figure}
\epsscale{1.25}
\vskip -0.05in
{\hskip  -0.15in \plotone{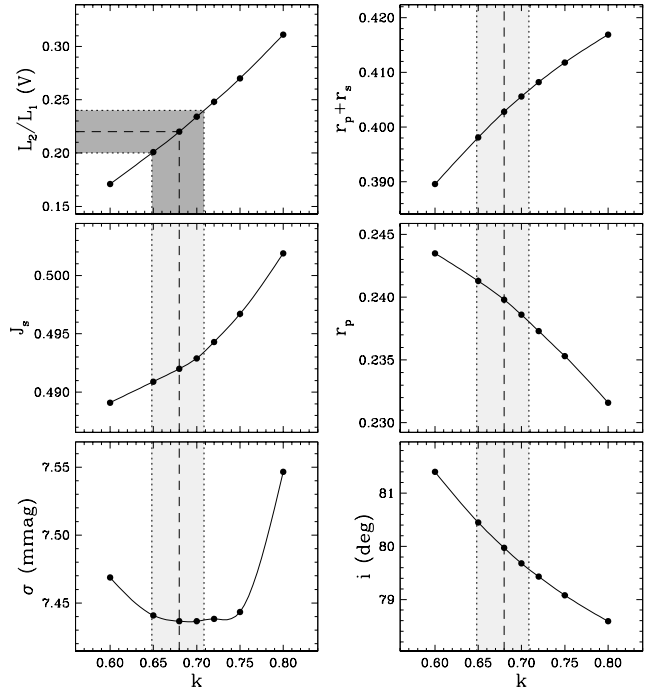}}
\vskip 0.2in
\figcaption[]{Application of the constraint given by the spectroscopic
brightness ratio to the light curve fits of VZ~Cep.  Grids of
solutions for fixed values of $k$ are shown for several key
parameters, along with the corresponding rms residual of the fit
($\sigma$).  The spectroscopic value $(L_{\rm B}/L_{\rm A})_V = 0.22
\pm 0.02$ is applied in the top left panel to determine $k$, and all
other quantities are interpolated to the same value.
\label{fig:constraint}}
\end{figure}

The $V$-band light ratio $(L_{\rm B}/L_{\rm A})_V$ from this fit is
consistent with our spectroscopic value from
\S\,\ref{sec:spectroscopy}, but formally less precise.  In similar
systems with partial eclipses, the \emph{accuracy} of the parameters
(more than their precision) is sometimes compromised because of strong
correlations among variables and the relatively flat bottom of the
$\chi^2$ surface in the least-squares problem \citep[see,
e.g.,][]{Andersen:91}.  In such systems it is often the case that more
accurate results are obtained by applying the spectroscopic light
ratio as an external constraint.  We have done this here by first
computing a grid of solutions for a range of fixed values of $k$.  We
then interpolated in the smooth relation obtained between the light
ratio and $k$ to our value of $(L_{\rm B}/L_{\rm A})_V = 0.22 \pm
0.02$, and derived $k = 0.680 \pm 0.030$. Interpolations of all other
quantities to this value of $k$ were then carried out.  This is
illustrated in Figure~\ref{fig:constraint} for some of the key
light-curve parameters.  Note that $\sigma$, the rms residual of the
fit, changes very little for $k$ between 0.65 and 0.75, demonstrating
that the radius ratio cannot be determined accurately from photometry
alone without an external constraint.  The results for the light curve
parameters from this constrained fit are listed in
Table~\ref{tab:ebopfits}, and are adopted for further use.  The
uncertainties have been propagated directly from the error in the
spectroscopic light ratio (see Figure~\ref{fig:constraint}), and
include also a contribution from the statistical uncertainties derived
from a separate solution in which $k$ was fixed at the best-fit value
and all other parameters were left free.

The fitted linear limb-darkening coefficients from this constrained
solution tend to be somewhat smaller than predicted by theory. We
find, for example, marginal differences of 0.10 ($\sim$1.3$\sigma$)
for both stars compared to the calculations by \cite{vanHamme:93}, and
more significant differences of 0.19 (2.5$\sigma$) compared to the
coefficients by \cite{Claret:00}. These differences are similar in
magnitude (and in this particular case, of the same sign) as those
reported, e.g., by \cite{Southworth:08}, and may be due to
shortcomings in the theoretical model atmospheres although
observational errors cannot be ruled out. For further comparisons
between theory and observations the reader is referred to the recent
work of \cite{Claret:08}. Adopting coefficients from the tables by
\cite{Claret:00} leads to values of the relative radii that are larger
by 1.1\% for the primary and 1.7\% for the secondary (1.6$\sigma$ and
0.5$\sigma$, respectively).

As a test of the reliability of the geometric parameters, we carried
out solutions with two other light-curve modeling programs that are
more sophisticated than the one we have used. One is the WINK program
\citep{Wood:72}, which adopts a better approximation to the stellar
shapes as triaxial ellipsoids, rather than the simpler biaxial
ellipsoids in EBOP, and includes a more detailed treatment of
reflection effects. The version we used has been modified and extended
as described by \cite{Vaz:84, Vaz:86}, \cite{Vaz:85}, and
\cite{Nordlund:90}. The other program is the Wilson-Devinney code
\citep[WD;][]{Wilson:71, Wilson:79, Wilson:90, Wilson:93, vanHamme:07}
in its most recent (2007) release, which uses Roche geometry. Light
curve solutions with these two codes were performed for a fixed value
of $k = 0.680$ (as closely as allowed by the different input
quantities) to permit a direct comparison with our constrained JKTEBOP
fit, and with the same limb-darkening law and coefficients as used
earlier. The WINK fit delivered marginally smaller relative radii that
differ from our JKTEBOP results by $\Delta r_{\rm A} = -0.0009$ and
$\Delta r_{\rm B} = -0.0005$ (i.e., less than 0.4\%), and an
inclination angle that was only $\Delta i = +0\fdg03$ larger. The WD
fit gave $\Delta r_{\rm A} = -0.0008$, $\Delta r_{\rm B} = -0.0005$,
and $\Delta i = +0\fdg22$. These results are thus not significantly
different from those of the simpler model we have used, as expected
from the relatively small ellipticity of the stars mentioned earlier.

The individual temperatures were determined from the central surface
brightness parameter $J_{\rm B}$ slightly adjusted for limb darkening
to correspond to the disk average \citep[see, e.g.,][]{Lacy:87}, the
absolute visual flux scale of \cite{Popper:80}, and an estimate of the
mean system temperature used as the initial value for the primary. The
latter was then improved by iteration until convergence.  The mean
system temperature is based on accurate Str\"omgren photometry for
VZ~Cep reported by \cite{Lacy:02a}.  Interstellar reddening was
estimated using the calibration of \cite{Perry:82} and the method of
\cite{Crawford:75}, which resulted in $E(b\!-\!y) = 0.032 \pm 0.007$
and an intrinsic color index of $(b\!-\!y)_0 = 0.286 \pm 0.007$.  The
calibration by \cite{Holmberg:07} was then used to derive a mean
system temperature of $6500 \pm 150$~K, assuming solar metallicity.
The individual temperatures derived in this way are $6690 \pm 160$~K
and $5720 \pm 120$~K for the primary and secondary, respectively,
which correspond to spectral types of approximately F3 and G4
\citep[][p.\ 430]{Gray:92}.  The primary $T_{\rm eff}$ is identical to
our spectroscopic estimate in \S\,\ref{sec:spectroscopy}.  The
temperature difference based the light curve is of course better
determined than the absolute values: $\Delta T_{\rm eff} = 970 \pm
35$~K.  Use of a different color/temperature calibration for inferring
the mean system temperature, such as that by \cite{Alonso:96}, yields
results only $\sim$30~K hotter.

\section{Absolute dimensions and physical properties} 
\label{sec:dimensions}

The spectroscopic and photometric solutions above lead to the masses
and radii given in Table~\ref{tab:absolute}.  Also included are the
predicted projected rotational velocities, calculated under the
assumption of synchronism with the orbital motion.  The secondary
value is consistent with our measured $v \sin i$ from
\S\,\ref{sec:spectroscopy}, but the expected value of $64.6 \pm
0.5$~\kms\ for the primary seems somewhat larger than our
spectroscopic estimate of $v \sin i = 57 \pm 3$~\kms.  At face value
this would indicate sub-synchronous rotation of that component, which
is unexpected in a short-period binary such as VZ~Cep.  Since the
primary star dominates the light of the system, we investigated the
possibility that there might be a photometric signal produced, for
instance, by rotational modulation from surface features on that
component. For this we examined the residuals from our adopted light
curve solution.  A Lomb-Scargle power spectrum did not indicate any
significant periodicities within a factor of two of the orbital
frequency, although the primary star is likely to be too hot for spots
to be important (see \S\,\ref{sec:discussion}).

There are no measurements of the chemical abundance of VZ~Cep. Our own
spectroscopy is inadequate for this, and the combined-light
Str\"omgren indices along with the calibration by \cite{Holmberg:07}
indicate [Fe/H] $= +0.06 \pm 0.09$, in which the uncertainties include
photometric errors as well as the scatter of the calibration. The {\it
Hipparcos\/} catalog \citep{Perryman:97} has no entry for VZ~Cep and
no trigonometric parallax is available. From its radiative properties
as measured here we find the distance to the system to be $215 \pm
10$~pc (similar distances are obtained separately for each component,
indicating the consistency of the measured properties).

Discrepancies described in the next section between our effective
temperature estimates and the $T_{\rm eff}$ values predicted by
stellar evolution models prompted us to attempt a deconvolution of the
combined-light photometry of VZ~Cep, as a check on both the color
excess and the temperature difference between the components.  We used
tables of standard Str\"omgren indices by \cite{Crawford:75} and
\cite{Olsen:84}, and synthesized binary stars for a range of primary
indices and a fixed $V$-band light ratio given by our spectroscopic
estimate of $(L_{\rm B}/L_{\rm A})_V = 0.22 \pm 0.02$.  We explored a
wide range of $E(b\!-\!y)$ values.  At each reddening we determined
the intrinsic indices for the primary and secondary that provide the
best match to the system values of $b\!-\!y$, $m_1$, $c_1$, and
$\beta$ as measured by \cite{Lacy:02a}, in the $\chi^2$ sense.  We
found the best agreement for $E(b\!-\!y) = 0.032$, in excellent accord
with our earlier determination based on the combined light.  The
measured $b\!-\!y$, $c_1$, and $\beta$ indices are reproduced to well
within their uncertainties, and $m_1$ is within 1.8$\sigma$. Making
use of the same color/temperature calibration by \cite{Holmberg:07}
invoked earlier, the intrinsic indices for each star from this
photometric deconvolution yield temperatures of 6670~K and 5690~K,
once again in very good agreement with the light curve results.  The
temperature difference from this exercise is $\Delta T_{\rm eff} = 975
\pm 40$~K.

\section{Comparison with stellar evolution theory}
\label{sec:evolution}

\begin{figure}
\epsscale{1.35}
\vskip -0.4in
{\hskip -0.35in \plotone{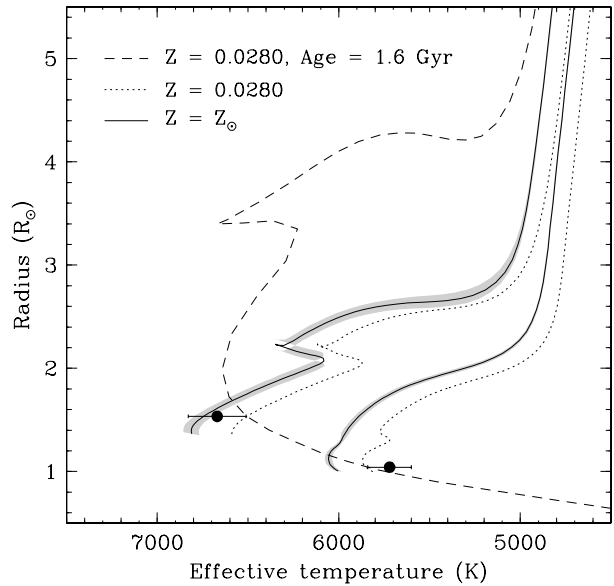}}
\vskip -0.3in
\figcaption[]{Stellar evolution models from the Yonsei-Yale series
\citep{Yi:01, Demarque:04} compared against the measurements for
VZ~Cep.  Solid lines show evolutionary tracks for the measured masses
and for solar metallicity ($Z = Z_{\sun}$), with the uncertainty in
the location of the tracks represented by the shaded regions. Dotted
lines correspond to mass tracks at a somewhat higher metallicity of $Z
= 0.0280$ that seems to fit the observations better (see text). An
isochrone for this metallicity and an age of 1.6~Gyr is shown with a
dashed curve. \label{fig:radteff}}
\end{figure}

The absolute masses of VZ~Cep have formal relative errors of 1\% or
better.  The radius of the primary is similarly well determined, while
that of the faint secondary is good to about 3.7\%.  These values
along with the temperatures are compared here with stellar evolution
models from the Yonsei-Yale series \citep{Yi:01, Demarque:04}.  In
Figure~\ref{fig:radteff} the measurements are shown in the $R$ vs.\
$T_{\rm eff}$ plane against evolutionary tracks computed for the
measured masses and for solar metallicity ($Z = 0.01812$ in these
models, indicated with solid lines).  The shaded areas represent the
uncertainty in the location of each track due to the measurement
errors in the masses $M_{\rm A}$ and $M_{\rm B}$.  While the primary
track is in good agreement with our temperature determination for that
star, the secondary track is too hot.  Adjustment of the chemical
composition of the models to $Z = 0.0280$ (corresponding to [Fe/H] $=
+0.21$) provides the fit shown with the dotted lines. This fit is
marginally consistent with our temperature error bars in the figure,
but the agreement is misleading since the temperature
\emph{difference} is much better determined than the error bars appear
to indicate. The best-fit age for this metallicity is 1.6~Gyr, and the
corresponding isochrone is indicated with a dashed curve.

\begin{figure}
\epsscale{1.35}
\vskip -0.4in
{\hskip -0.35in \plotone{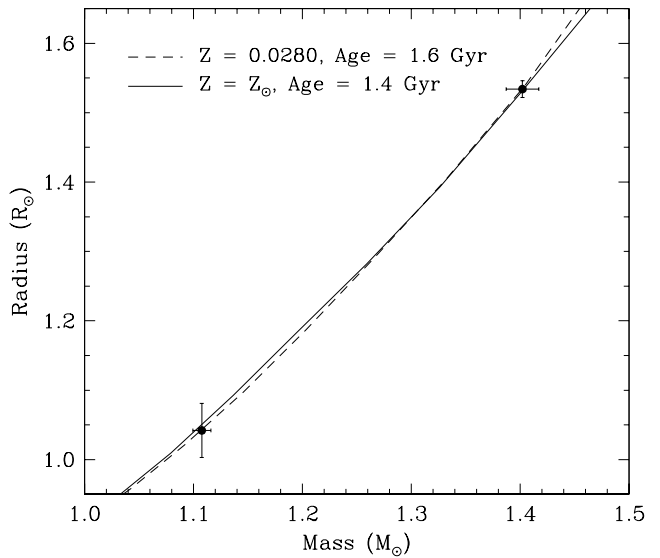}}
\vskip -0.3in
\figcaption[]{Isochrones from the Yonsei-Yale series \citep{Yi:01,
Demarque:04} compared with the measurements for VZ~Cep in the
mass-radius plane. The dashed line is the same isochrone shown in
Figure~\ref{fig:radteff} ($Z = 0.0280$), and a solar-metallicity
isochrone is represented by the solid curve, for a slightly younger
age that fits the observations best. \label{fig:massrad}}
\end{figure}

Figure~\ref{fig:massrad} shows the measurements in the mass-radius
diagram against the same set of models. The dashed line represents the
same isochrone shown before, and the solid line is an isochrone for
solar metallicity that provides the best fit, in this case for a
slightly younger age of 1.4~Gyr. Both are seen to represent the
measurements equally well.

These comparisons suggest that the models correctly predict the radii
of the stars at the measured masses, but that the temperature of the
secondary is underestimated by a significant amount. Tests with a
different series of models by \cite{Pietrinferni:04} gave similar
results.

\section{Discussion and conclusions} 
\label{sec:discussion}

VZ~Cep stands out among the F stars as one of the eclipsing binaries
with the largest difference in mass between the components ($q =
0.7900$), which provides extra leverage for testing stellar evolution
models. There are no less than five other systems with well determined
properties \citep[BW~Aqr~B, V1143~Cyg~A, BP~Vul~B, V442~Cyg~B, and
AD~Boo~A;][]{Andersen:91, Lacy:03, Clausen:08} that have at least one
component nearly identical in mass to the primary of VZ~Cep (i.e.,
within 1\%).  However, these stars are all in very different
evolutionary states so that their radii span a range of 33\% and their
effective temperatures differ by up to 360~K. They are therefore of
little help in understanding the discrepancies with theory noted above
for VZ~Cep.  Only one other well measured binary has one component
with a mass similar to that of the secondary of VZ~Cep, but that star
(EK~Cep~B) is considered to be in the pre-main sequence stage
\citep{Popper:87}.

Figure~\ref{fig:radteff} highlights the main disagreement between the
models and the measurements for VZ~Cep, which is that the temperature
difference predicted from theory for the measured masses and surface
gravities is much smaller than all of our estimates. Solar metallicity
models, which appear to fit the properties of the primary well,
indicate $\Delta T_{\rm eff} = 710$~K, and this is reduced further to
660~K when considering the higher metallicity of $Z = 0.0280$. The
uncertainty in these determinations is difficult to quantify, but a
useful measure may be obtained by propagating the uncertainty in the
measured masses, which results in $\pm$50~K.  Uncertainties from
physical inputs to the models are unlikely to add much to this due to
the differential nature of the comparison.  In this paper we have made
three empirical determinations of $\Delta T_{\rm eff}$, as follows: 1)
$\Delta T_{\rm eff} = 970 \pm 35$~K, based on the $J_{\rm B}$ value
from our light-curve analysis along with the visual flux scale of
\cite{Popper:80} and our spectroscopic brightness ratio (used as an
external constraint); 2) $\Delta T_{\rm eff} = 975 \pm 40$~K, from
photometric deconvolution based on the measured Str\"omgren indices
and the spectroscopic brightness ratio (\S\,\ref{sec:dimensions}),
along with the color/temperature calibrations of \cite{Holmberg:07};
3) $\Delta T_{\rm eff} = 940$~K, directly from a primary temperature
estimate based on spectroscopy (\S\,\ref{sec:spectroscopy}) and an
assumed temperature for the secondary similar to estimates for that
star from the other two methods. While these three determinations are
not completely independent, their consistency despite the widely
different ingredients reinforces our conclusion that the model $\Delta
T_{\rm eff}$ is at least $\sim$250~K too small.

Stellar evolution models have been shown previously to overestimate
the effective temperatures of low-mass stars in eclipsing binaries by
up to $\sim$200~K \citep[e.g.,][]{TorresRibas:02, Ribas:03}.  The
study of V1061~Cyg by \cite{Torres:06} suggested that the problem is
not confined to M dwarfs, but extends up to masses almost as large as
that of the Sun (0.93~M$_{\sun}$ in the case of V1061~Cyg~B). At the
same time, the radii of these stars appear too large compared to
theory, and both discrepancies are generally attributed to the effects
of stellar activity in these short-period, tidally synchronized and
rapidly rotating systems.  

There is little doubt that the VZ~Cep system is active, judging by its
strong X-ray emission as recorded by ROSAT \citep{Voges:99}. We
estimate its X-ray luminosity to be $\log L_{\rm X} = 30.61 \pm 0.12$
(where $L_{\rm X}$ is in cgs units), and $\log L_{\rm X}/L_{\rm bol} =
-3.70 \pm 0.13$.\footnote{By comparison, $\log L_{\rm X}$ for the Sun
ranges between 26.4 and 27.7 during the activity cycle
\citep{Peres:00}, and $\log L_{\rm X}/L_{\rm bol}$ ranges between
$-7.2$ and $-5.9$.}  The mass of VZ~Cep~B is slightly larger than that
of the Sun, but it still has a thin convective envelope representing
about 1.3\% of the total mass, which suggests that star may in fact be
responsible for most of the X-ray emission given that the primary has
no convective envelope. Another indication is given by the Rossby
numbers of the stars (ratio $R_0$ between the convective turnover time
and the rotational period). For the primary we estimate $\log R_0 >
2.1$, which according to \citet[Fig.~6]{Hall:94} clearly places it
among the inactive stars. The secondary, on the other hand, has $\log
R_0 = -1.3$. Stars in this regime tend to be very active and have
photometric variability due to spots with amplitudes as large as
$\sim$0.4 mag.  Detection of this expected variability is difficult in
VZ~Cep because of the faintness of the secondary. Nevertheless, we
examined the nightly residuals from our adopted solution near the
primary minimum, where the contrast is more favorable, and we see only
occasional systematic deviations on one or two nights. However,
similar deviations are seen at the secondary eclipse, and also outside
of eclipse, which leads us to believe these are residual instrumental
errors rather than real changes in the light level caused by
spottedness on the secondary (see \S\,\ref{sec:photometry}).

If we consider the properties of the primary of VZ~Cep to be
relatively well described by theory for a metallicity near solar, then
the secondary shows a temperature difference compared to models in the
same direction as mentioned above for other active stars (i.e., lower
than predicted).  However, we see no indication that its radius is
larger than predicted (Figure~\ref{fig:massrad}), which we would have
expected not only from the evidence displayed by other systems but
also from recent theoretical studies of the effects of chromospheric
activity \citep[e.g.,][]{Mullan:01, Chabrier:07}.

As described in previous sections, we have carried out a variety of
tests to explore the possibility of systematic errors in our mass,
radius, or temperature determinations, including a careful examination
of biases in our velocity measurements, and sanity checks of our
light-curve analysis with results from more sophisticated modeling
programs. We were unable to demonstrate any significant errors that
would explain the discrepancies in the preceding paragraph.  For
example, matching the model $\Delta T_{\rm eff}$ with the $\Delta
T_{\rm eff}$ measured from the light curve would require a change in
the mass ratio to $q \approx 0.71$, much lower than allowed by the
spectroscopy, regardless of the choice of cross-correlation templates
(see \S\,\ref{sec:spectroscopy}).  Conversely, deriving a smaller
temperature difference from the light curve to match the models would
require an increase in $J_{\rm B}$ to values that are unrealistic and
would bring strong disagreement with the light ratio from
spectroscopy. Additionally, this would leave the other two empirical
estimates of $\Delta T_{\rm eff}$ unchanged, and a discrepancy would
remain.  As indicated earlier, we see no evidence for third light at a
level that would make much difference. The adjustments required in
each of the quantities mentioned above, and others we experimented
with, are so large compared to the uncertainties that a combination of
effects is unlikely to resolve the issue either.

\begin{figure}
\epsscale{1.35}
\vskip -0.4in
{\hskip -0.35in \plotone{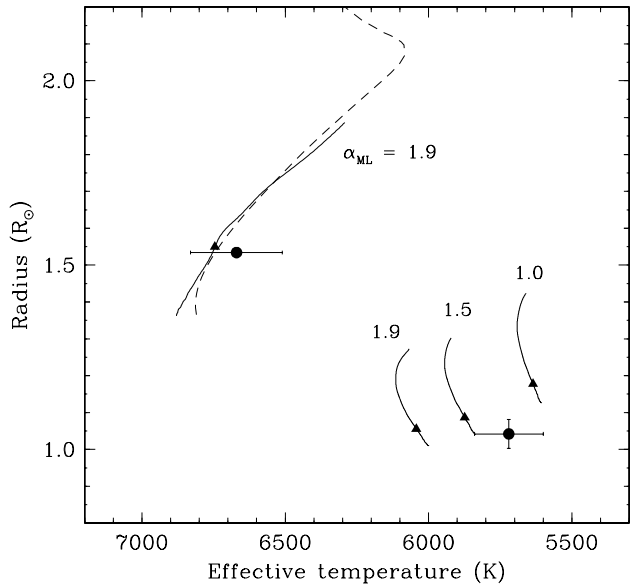}}
\vskip -0.3in
\figcaption[]{Radius and effective temperature of VZ~Cep compared
against evolutionary tracks for solar metallicity by
\cite{Baraffe:98}, for the exact masses we measure (solid lines). A
single track is shown for the primary star, for a mixing length
parameter $\alpha_{\rm ML} = 1.9$ appropriate for the Sun. The dashed
line represents the same solar-metallicity Yonsei-Yale track shown in
Figure~\ref{fig:radteff}. Three \cite{Baraffe:98} models are shown for
the secondary, for different values of $\alpha_{\rm ML}$, as
labeled. For reference, the triangles on the solid curves correspond
to an age of 1.4~Gyr, which was found in Figure~\ref{fig:massrad} to
provide a good match in the mass-radius plane using solar-metallicity
models from the Yonsei-Yale series.
\label{fig:baraffe}}
\end{figure}

At the suggestion of the referee, we show in Figure~\ref{fig:baraffe}
a comparison with an alternate set of models by \cite{Baraffe:98},
which allows us to explore the effect of differences in the mixing
length parameter $\alpha_{\rm ML}$. Previous studies of the radius and
temperature discrepancies for active low-mass stars have indicated
that a value of $\alpha_{\rm ML}$ lower than appropriate for the Sun,
representing a reduced overall convective efficiency, provides a much
better match to the observations of these objects.  Consequently, we
show a solar-metallicity evolutionary track for a solar-like mixing
length parameter for the radiative primary ($\alpha_{\rm ML} = 1.9$ in
these models), and tracks for three values of the mixing length
parameter for the secondary star ($\alpha_{\rm ML} = 1.9$, 1.5, and
1.0), which, as mentioned earlier, we believe to be the more active
member of the system. For reference, triangles on each track mark the
age of 1.4~Gyr, which we found in Figure~\ref{fig:radteff} to provide
the best match for $Z = Z_{\sun}$ using the Yonsei-Yale models.  The
\cite{Baraffe:98} model for the primary is seen to be similar to the
corresponding solar-metallicity Yonsei-Yale model (dashed line,
reproduced from Figure~\ref{fig:radteff}). A reduction of the mixing
length parameter for the secondary star leads to the expected
systematic decrease in effective temperature, and an increase in
radius. A secondary model with $\alpha_{\rm ML}$ between 1.0 and 1.5,
when paired with the standard $\alpha_{\rm ML} = 1.9$ model for the
primary, would appear to give approximately the correct temperature
difference for the system. However, the measured radius of VZ~Cep~B is
somewhat smaller than predicted, in agreement with our earlier
conclusion that this star appears normal in size (compared to standard
models).

The 13\% difference between our measured $v \sin i$ for the primary
and the predicted synchronous velocity is somewhat puzzling, and is
significant at the 2.5-$\sigma$ confidence level. We do not believe
errors in the spectroscopic measurements are to blame since all 39 of
our individual spectra consistently give values smaller than
predicted.  A reduction in the predicted value could be accomplished
with an increase in $k$, but it would have to be much larger than
allowed by our photometric solutions, and would once again bring
disagreement between the photometric and spectroscopic light ratios.
	
At the moment we are unable to offer an explanation for the
differences noted above, and based on the tests just described we are
inclined to believe that the measurements are accurate and that the
system is affected in some way that the models do not account for,
most likely having to do with chromospheric activity. It would also
appear that our understanding of the effects of chromospheric activity
(reduced convective efficiency, spot coverage) on the global
properties of stars is still incomplete, since we see here only the
effect on the temperature predicted by recent models that account for
these phenomena \citep{Chabrier:07}, but not the effect on the radius.
VZ~Cep thus presents a challenge to theory. Further progress in
understanding these differences may be made by obtaining complete
light curves in multiple passbands, which would give a better handle
on the temperature issue. Higher signal-to-noise ratio spectroscopy
would also help in refining the $v \sin i$ measurements for the
primary and secondary, in constraining the abundance, and perhaps also
in providing a more direct determination of the effective temperatures
and revealing whether \ion{Ca}{2} emission is present.

\acknowledgments

The spectroscopic observations of VZ~Cep used in this paper were
obtained with the expert assistance of P.\ Berlind, M.\ Calkins, D.\
W.\ Latham, and R.\ P.\ Stefanik. R.\ J.\ Davis is thanked for
maintaining the CfA echelle database. We are grateful to the referee,
J.\ V.\ Clausen, for a prompt, detailed, and very helpful report. GT
acknowledges partial support for this work from NSF grant AST-0708229.
CHSL would like to thank University of Arkansas graduate student
Kathryn D.\ Hicks for a preliminary analysis of the photometry and
radial velocities of VZ~Cep.  This research has made use of the SIMBAD
database, operated at CDS, Strasbourg, France, and of NASA's
Astrophysics Data System Abstract Service.

\clearpage

\begin{deluxetable}{lcccc}
\tabletypesize{\scriptsize}
\tablewidth{0pc}
\tablecaption{Published measurements of the times of eclipse for
VZ~Cep.\label{tab:minima}}
\tablehead{
\colhead{HJD} &
\colhead{} &
\colhead{Uncertainty} &
\colhead{$(O\!-\!C)$} &
\colhead{} \\
\colhead{\hbox{~~(2,400,000$+$)~~}} &
\colhead{Type\tablenotemark{a}} &
\colhead{(days)} &
\colhead{(days)} &
\colhead{Source}
}
\startdata
 49567.42210\dotfill &  1  &  0.0003\phn  &  $+$0.00067 &  1 \\
 51608.72370\dotfill &  1  &  0.0003\phn  &  $-$0.00023 &  2 \\
 52038.87680\dotfill &  2  &  0.0005\phn  &  $-$0.00012 &  3 \\
 52044.79410\dotfill &  2  &  0.0005\phn  &  $+$0.00036 &  3 \\
 52051.89410\dotfill &  2  &  0.0005\phn  &  $+$0.00018 &  3 \\
 52054.85215\dotfill &  1  &  0.00011 &  $+$0.00008 &  3 \\
 52073.78570\dotfill &  1  &  0.0002\phn  &  $-$0.00019 &  3 \\
 52076.74440\dotfill &  2  &  0.0003\phn  &  $-$0.00016 &  3 \\
 52079.70300\dotfill &  1  &  0.0003\phn  &  $+$0.00029 &  3 \\
 52080.88604\dotfill &  1  &  0.00019 &  $-$0.00003 &  3 \\
 52093.90270\dotfill &  1  &  0.0005\phn  &  $-$0.00038 &  3 \\
 52108.69630\dotfill &  2  &  0.0003\phn  &  $+$0.00092 &  3 \\
 52111.65270\dotfill &  1  &  0.0006\phn  &  $-$0.00083 &  3 \\
 52112.83709\dotfill &  1  &  0.00014 &  $+$0.00019 &  3 \\
 52114.61500\dotfill &  2  &  0.0006\phn  &  $+$0.00280 &  3 \\
 52154.84470\dotfill &  2  &  0.0004\phn  &  $-$0.00187 &  3 \\
 52159.58000\dotfill &  2  &  0.0010\phn  &  $-$0.00002 &  3 \\
 52166.67970\dotfill &  2  &  0.0003\phn  &  $-$0.00051 &  3 \\
 52179.69710\dotfill &  2  &  0.0003\phn  &  $-$0.00011 &  3 \\
 52233.54070\dotfill &  1  &  0.0004\phn  &  $+$0.00070 &  3 \\
 52277.32429\dotfill &  1  &  0.00007 &  $-$0.00017 &  4 \\
 52463.70530\dotfill &  2  &  0.0003\phn  &  $+$0.00079 &  5 \\
 52464.88810\dotfill &  2  &  0.0003\phn  &  $+$0.00022 &  5 \\
 52482.63870\dotfill &  2  &  0.0005\phn  &  $+$0.00037 &  5 \\
 52518.73064\dotfill &  1  &  0.00019 &  $-$0.00003 &  5 \\
 53366.01950\dotfill &  1  &  0.0002\phn  &  $+$0.00037 &  6 \\
 53658.30900\dotfill &  1  &  0.0006\phn  &  $-$0.00098 &  7 \\
 54009.76910\dotfill &  1  &  0.0001\phn  &  $+$0.00008 &  8 \\ [-1.5ex]
\enddata
\tablenotetext{a}{Type: 1 = primary eclipse; 2 = secondary eclipse.}
\tablecomments{References: 1. \cite{Agerer:95}; 2. \cite{Nelson:01};
3. \cite{Lacyetal:02}; 4. \cite{Sarounova:05}; 5. \cite{Lacy:02b};
6. \cite{Kim:06}; 7. \cite{Diethlem:06}; 8. \cite{Nelson:07}.}
\end{deluxetable}

\clearpage

\begin{deluxetable}{lccccc}
\tabletypesize{\scriptsize}
\tablewidth{0pc}
\tablecaption{Radial velocity measurements of VZ~Cep.\label{tab:rvs}}
\tablehead{\colhead{HJD} &
\colhead{$RV_{\rm A}$} &
\colhead{$RV_{\rm B}$} &
\colhead{$(O\!-\!C)_{\rm A}$} &
\colhead{$(O\!-\!C)_{\rm B}$} &
\colhead{} \\
\colhead{\hbox{~~(2,400,000$+$)~~}} &
\colhead{(\kms)} &
\colhead{(\kms)} &
\colhead{(\kms)} &
\colhead{(\kms)} &
\colhead{Orbital phase}}
\startdata
  52661.5709\dotfill &  $+$106.35 &  $-$149.88  &  $+$1.69 &  $+$2.71 &   0.707 \\
  52808.9615\dotfill &  $-$127.46 &  $+$139.25  &  $+$1.12 &  $-$3.40 &   0.259 \\
  52828.9269\dotfill &   \phn$-$96.88 &  $+$102.39  &  $+$0.15 &  $-$0.32 &   0.131 \\
  52834.9798\dotfill &  $-$127.64 &  $+$147.21  &  $+$1.10 &  $+$4.36 &   0.246 \\
  52885.8243\dotfill &  $-$126.22 &  $+$134.87  &  $-$0.81 &  $-$3.76 &   0.212 \\
  52894.8309\dotfill &   \phn$+$96.39 &  $-$136.82  &  $-$0.31 &  $+$5.69 &   0.823 \\
  52924.7873\dotfill &  $-$100.02 &  $+$112.30  &  $+$0.34 &  $+$5.38 &   0.138 \\
  52951.7261\dotfill &   \phn$+$57.66 &  $-$102.06  &  $-$0.97 &  $-$7.73 &   0.902 \\
  52958.6957\dotfill &  $+$105.49 &  $-$155.58  &  $+$0.59 &  $-$2.69 &   0.792 \\
  53013.5806\dotfill &  $-$114.11 &  $+$129.47  &  $+$0.78 &  $+$4.16 &   0.172 \\
  53182.9448\dotfill &  $-$125.76 &  $+$138.43  &  $-$1.35 &  $+$1.06 &   0.293 \\
  53185.9637\dotfill &   \phn$+$88.37 &  $-$131.20  &  $-$0.31 &  $+$1.16 &   0.845 \\
  53186.9164\dotfill &   \phn$+$88.42 &  $-$133.24  &  $+$2.37 &  $-$4.21 &   0.650 \\
  53189.9839\dotfill &  $-$129.03 &  $+$137.70  &  $-$0.41 &  $-$4.99 &   0.242 \\
  53191.8776\dotfill &   \phn$+$89.70 &  $-$128.04  &  \phs0.00 &  $+$5.61 &   0.842 \\
  53192.9211\dotfill &  $+$106.23 &  $-$154.75  &  $-$1.14 &  $+$1.27 &   0.724 \\
  53215.9246\dotfill &  $-$112.11 &  $+$119.86  &  $-$0.73 &  $-$1.02 &   0.163 \\
  53218.8991\dotfill &   \phn$+$94.48 &  $-$137.57  &  $-$2.02 &  $+$4.69 &   0.677 \\
  53271.6999\dotfill &  $-$122.28 &  $+$137.26  &  $+$1.64 &  $+$0.51 &   0.296 \\
  53272.7658\dotfill &  $-$122.11 &  $+$134.75  &  $-$0.01 &  $+$0.31 &   0.197 \\
  53274.6998\dotfill &   \phn$+$94.19 &  $-$144.23  &  $+$0.17 &  $-$5.11 &   0.831 \\
  53275.7597\dotfill &  $+$104.75 &  $-$162.66  &  $-$2.93 &  $-$6.25 &   0.727 \\
  53281.7191\dotfill &  $+$107.50 &  $-$158.17  &  $-$1.13 &  $-$0.56 &   0.763 \\
  53282.8175\dotfill &  $+$100.85 &  $-$144.52  &  $+$0.05 &  $+$3.18 &   0.691 \\
  53301.7530\dotfill &  $+$100.90 &  $-$147.17  &  $-$0.28 &  $+$1.01 &   0.692 \\
  53333.7365\dotfill &  $+$106.56 &  $-$150.44  &  $-$0.27 &  $+$4.89 &   0.720 \\
  53335.6589\dotfill &  $-$109.05 &  $+$117.57  &  $-$0.46 &  $+$0.23 &   0.344 \\
  53336.6417\dotfill &  $-$116.21 &  $+$124.74  &  $-$0.51 &  $-$1.60 &   0.175 \\
  53339.6764\dotfill &  $+$108.41 &  $-$157.55  &  $-$0.30 &  $+$0.16 &   0.739 \\
  53630.7663\dotfill &  $+$106.36 &  $-$156.85  &  $-$1.07 &  $-$0.75 &   0.724 \\
  53636.7401\dotfill &  $+$108.74 &  $-$160.82  &  $+$0.93 &  $-$4.24 &   0.773 \\
  53690.6534\dotfill &  $-$112.94 &  $+$127.71  &  $+$0.49 &  $+$4.24 &   0.332 \\
  53691.7072\dotfill &  $-$127.60 &  $+$134.58  &  $-$0.62 &  $-$6.04 &   0.222 \\
  54042.6132\dotfill &  $+$111.73 &  $-$160.65  &  $+$2.80 &  $-$2.66 &   0.755 \\
  54043.7068\dotfill &   \phn$+$98.06 &  $-$145.42  &  $+$0.68 &  $-$2.05 &   0.679 \\
  54074.5790\dotfill &  $+$106.66 &  $-$154.30  &  $-$1.60 &  $+$2.85 &   0.768 \\
  54103.5713\dotfill &  $-$127.62 &  $+$139.12  &  $+$0.44 &  $-$2.87 &   0.268 \\
  54279.9427\dotfill &  $-$121.65 &  $+$138.43  &  $-$1.19 &  $+$6.07 &   0.310 \\
  54282.8950\dotfill &  $+$105.11 &  $-$148.54  &  $+$3.09 &  $+$0.71 &   0.805 \\ [-1.5ex]
\enddata
\tablecomments{These velocities include corrections for systematics (see text).}
\end{deluxetable}

\clearpage

\begin{deluxetable}{lc}
\tablewidth{0pc}
\tablecaption{Spectroscopic orbital solution for VZ~Cep.\label{tab:specorbit}}
\tablehead{
\colhead{
\hfil~~~~~~~~~~~~~Parameter~~~~~~~~~~~~~~} & \colhead{Value}}
\startdata
\multicolumn{2}{l}{Adjusted quantities\hfil} \\
~~~~$P$ (days)\tablenotemark{a}\dotfill        &  1.18336377                   \\
~~~~$T_{\rm I}$ (HJD$-$2,400,000)\tablenotemark{a}\dotfill     & 52,277.32446  \\
~~~~$K_{\rm A}$ (\kms)\dotfill                 &  118.88~$\pm$~0.22\phn\phn    \\
~~~~$K_{\rm B}$ (\kms)\dotfill                 &  150.48~$\pm$~0.67\phn\phn    \\
~~~~$\gamma$ (\kms)\dotfill                    & $-9.90$~$\pm$~0.21\phs        \\
~~~~$\Delta RV$ (\kms)\tablenotemark{b}\dotfill   &  $-2.31$~$\pm$~0.65\phs    \\
\multicolumn{2}{l}{Derived quantities\hfil} \\
~~~~$M_{\rm A}\sin^3 i$ (M$_{\sun}$)\dotfill &  1.339~$\pm$~0.013              \\
~~~~$M_{\rm B}\sin^3 i$ (M$_{\sun}$)\dotfill &  1.0577~$\pm$~0.0064            \\
~~~~$q\equiv M_{\rm B}/M_{\rm A}$\dotfill    &  0.7900~$\pm$~0.0038            \\
~~~~$a_{\rm A}\sin i$ (10$^6$ km)\dotfill    &  1.9345~$\pm$~0.0036            \\
~~~~$a_{\rm B}\sin i$ (10$^6$ km)\dotfill    &  2.4487~$\pm$~0.0109            \\
~~~~$a \sin i$ (R$_{\sun}$)\dotfill          &  6.298~$\pm$~0.016              \\
\multicolumn{2}{l}{Other quantities pertaining to the fit\hfil} \\
~~~~$N_{\rm obs}$\dotfill                    & 39                              \\
~~~~Time span (days)\dotfill                 &  1621.3                         \\
~~~~$\sigma_{\rm A}$ (\kms)\dotfill          & 1.27                            \\
~~~~$\sigma_{\rm B}$ (\kms)\dotfill          & 3.82                            \\ [-1.0ex]
\enddata
\tablenotetext{a}{Ephemeris adopted from \S\,\ref{sec:ephemeris}.}
\tablenotetext{b}{Velocity offset in the sense
$\langle$primary$-$secondary$\rangle$ (see text).}
\end{deluxetable}


\begin{deluxetable}{ccc}
\tablewidth{0pc}
\tablecaption{Differential $V$-band magnitudes of VZ~Cep.\label{tab:photometry}}
\tablehead{
\colhead{HJD$-2,\!440,\!000$} &
\colhead{$\Delta V$} &
\colhead{Orbital phase}}
\startdata
 51973.95823  &  0.129  &   0.64077  \\
 51973.95926  &  0.115  &   0.64164  \\
 51973.96028  &  0.123  &   0.64250  \\
 51973.96130  &  0.126  &   0.64336  \\
 51973.96233  &  0.126  &   0.64423  \\ [-1.5ex]
\enddata
\tablecomments{Table~\ref{tab:photometry} is available in its entirety
in the electronic edition of the {\it Astronomical Journal}.  A
portion is shown here for guidance regarding its form and contents.}
\end{deluxetable}


\begin{deluxetable}{lcc}
\tablewidth{0pc}
\tablecaption{Photometric orbital solutions for VZ~Cep.\label{tab:ebopfits}}
\tablehead{
\colhead{~~~~Parameter~~~~} &
\colhead{Unconstrained fit} &
\colhead{Constrained fit}}
\startdata
$J_{\rm B}$\dotfill     &      0.495~$\pm$~0.010  &   0.4920~$\pm$~0.0013 \\
$k \equiv r_{\rm B}/r_{\rm A}$\dotfill &   0.717~$\pm$~0.041  &   0.680~$\pm$~0.030 \\
$r_{\rm A}+r_{\rm B}$\dotfill  & 0.4077~$\pm$~0.0043  &   0.4028~$\pm$~0.0077 \\
$r_{\rm A}$\dotfill     &      0.2374~$\pm$~0.0033 &  0.2398~$\pm$~0.0017 \\
$r_{\rm B}$\dotfill     &      0.1703~$\pm$~0.0073  &   0.1630~$\pm$~0.0061 \\
$u_{\rm A}$\dotfill     &      0.499~$\pm$~0.075  &     0.420~$\pm$~0.076 \\
$u_{\rm B}$\dotfill     &      0.581~$\pm$~0.059  &     0.500~$\pm$~0.076 \\
$i$ (deg)\dotfill       &       79.47~$\pm$~0.44\phn   &     79.97~$\pm$~0.45\phn \\
$L_{\rm A}(V)$\tablenotemark{a}\dotfill     &      0.802~$\pm$~0.022  &   0.820~$\pm$~0.014 \\
$(L_{\rm B}/L_{\rm A})_V$\dotfill   &     0.246~$\pm$~0.033  &   0.220~$\pm$~0.020\tablenotemark{b} \\
$\sigma_V$ (mmag)\dotfill  &    7.4400       &        7.4365 \\
$N_{\rm obs}$\dotfill   &         5473      &         5473 \\ [-1.0ex]
\enddata
\tablenotetext{a}{Fractional luminosity of the primary.}
\tablenotetext{b}{Adopted as a constraint from spectroscopy (see text).}
\end{deluxetable}

\clearpage

\begin{deluxetable}{lcc}
\tablewidth{0pc}
\tablecaption{Physical properties of VZ~Cep.\label{tab:absolute}}
\tablehead{
\colhead{~~~~~~~Parameter~~~~~~~} &
\colhead{Primary} &
\colhead{Secondary}}
\startdata
Mass (M$_{\sun}$)\dotfill            &   1.402~$\pm$~0.015   &     1.1077~$\pm$~0.0083 \\
Radius (R$_{\sun}$)\dotfill          &   1.534~$\pm$~0.012   &     1.042~$\pm$~0.039 \\
$\log g$ (cgs)\dotfill               &   4.2130~$\pm$~0.0080 &     4.446~$\pm$~0.033 \\
Temperature (K)\dotfill              &   6670~$\pm$~160\phn  &     5720~$\pm$~120\phn \\
$\log L$ (L$_{\sun}$)\dotfill        &   0.634~$\pm$~0.041   &     0.026~$\pm$~0.050 \\
$v \sin i$ (\kms)\tablenotemark{a}\dotfill            &   57~$\pm$~3\phn      &     50~$\pm$~10   \\
$v_{\rm sync} \sin i$ (\kms)\tablenotemark{b}\dotfill &   64.6~$\pm$~0.5\phn  &     43.9~$\pm$~1.6\phn \\
$a$ (R$_{\sun}$)\dotfill             &  \multicolumn{2}{c}{6.396~$\pm$~0.019} \\
Distance (pc)\dotfill                &  \multicolumn{2}{c}{215~$\pm$~10\phn} \\
$M_{\rm bol}$ (mag)\dotfill          &   3.18~$\pm$~0.10     &     4.68~$\pm$~0.12 \\
$M_V$ (mag)\dotfill                  &   3.16~$\pm$~0.11     &     4.77~$\pm$~0.12 \\ [-1.5ex]
\enddata
\tablenotetext{a}{Value measured spectroscopically.}
\tablenotetext{b}{Value predicted assuming synchronous rotation.}
\end{deluxetable}

\end{document}